\documentclass[twocolumn,apj]{emulateapj-rtx4}

\usepackage{natbib}
\usepackage{graphicx}
\usepackage{amsmath,amssymb}
\usepackage{graphicx}
\usepackage{txfonts}
\usepackage{epstopdf}

\usepackage{ulem} 
\usepackage{natbib}
\newcommand{\ord}[1]{{{\cal O}({#1})}}
\newcommand{\vect}[1]{{\boldsymbol{#1}}}

\newcommand{\SF}{\mathrm{SF}}
\newcommand{\kk}{\mathrm{k}}

\begin{document}
\title{Imprints of expansion onto the local anisotropy of solar wind turbulence}
\author{Andrea~Verdini\altaffilmark{1}}
\affil{Dipartimento di fisica e astronomia, Universit\`a di Firenze,
Firenze, Italy.}
\author{Roland~Grappin}
\affil{LPP, Ecole Polytechnique, Palaiseau, France.}
\altaffiltext{1}{Solar-Terrestrial Center of Excellence - SIDC, Royal
Observatory of Belgium, Bruxelles, Belgium.}

 \date{\today}
 
\begin{abstract}
We study the anisotropy of II-order structure functions defined in a frame
attached to the local mean field in three-dimensional (3D) direct numerical
simulations of magnetohydrodynamic turbulence, including or not the solar wind expansion.
We simulate spacecraft flybys through the numerical domain by taking increments
along the radial (wind) direction that forms an angle of $45^o$ with the
ambient magnetic field.
We find that only when expansion is taken into account, do the synthetic
observations match the 3D anisotropy observed in the solar wind,
including the change of anisotropy with scales.
Our simulations also show that the anisotropy  changes dramatically when
considering increments oblique to the radial directions. 
Both results can be understood by noting that expansion reduces the
radial component of the magnetic field at all scales, thus confining fluctuations in the plane perpendicular to the radial.
Expansion is thus shown to affect not only the (global) spectral anisotropy, but
also the local anisotropy of second-order structure functions by
influencing the distribution of the local mean field, which 
enters this higher-order statistics.
\end{abstract}
   \maketitle
\section{Introduction}
The solar wind is known to be a turbulent medium since many decades \citep{Coleman_1968} and is probably the best example of natural turbulent laboratory in astrophysics \citep[e.g.][]{Bruno_Carbone_2013}. 
Turbulence shows most of the time a non zero global mean field $\vect{B_0}$,
which should lead to an anisotropic cascade with the spectrum being
axisymmetric around it
\citep{Montgomery_Turner_1981,Shebalin_al_1983,Grappin_1986}.
As the angle between $\vect{B_0}$ and the radial direction varies,
a spacecraft embedded in the radial solar wind samples 
data in different directions with respect to the mean field. This
allows one to measure the correlation function in two dimensions, which 
has the chacteristic of an anisotropic cascade \citep{Matthaeus_al_1990,Bieber_al_1996,Dasso_al_2005,Hamilton_al_2008}.

However, the axisymmetry assumption has been found to break down in several
works \citep{Saur_Bieber_1999,Narita_al_2010,Chen_al_2012}.
This may result from (a), considering scales large enough for the expansion to
play a role and/or (b), considering anisotropy with
respect to the local mean field instead of the global mean field.
While having a reference frame attached to the global $\vect{B_0}$ is
preferable for studying the turbulent dissipation
\citep[e.g.][]{Verdini_al_2015}, a reference frame attached to the local,
scale-dependent, mean field ($\vect{B_\ell}$) allows one to reveal the effect
of local dynamics in magnetohydrodynamic (MHD) turbulence.
In the latter case a different scaling in the two directions parallel and
perpendicular to $\vect{B_\ell}$ was found both in direct numerical
simulations (DNS) \citep[e.g.][]{Cho_Vishniac_2001,Milano_al_2001,
Beresnyak_2009,Grappin_al_2013}
and in the solar wind \citep[e.g.][]{Horbury_al_2008,
Podesta_2009,Luo_Wu_2010,Wicks_al_2010,Wicks_al_2011,Wicks_al_2012,Wicks_al_2013,Chen_al_2011,He_al_2013}.

Deviations from axisymmetry (in the form of three distinct scaling laws) appear
when considering two perpendicular directions
instead of a single one \citep[see ][]{Boldyrev_2006}.
In their recent measurements \citet{Chen_al_2012} show how the small scale
ordering of the structure functions (SF) is completely modified in the solar wind when passing from small to large scales.
While the small-scale anisotropy is roughly compatible with
three-dimensional anisotropic phenomenology of turbulence \citep{Boldyrev_2006},
the large-scale anisotropy has no explanation so far.

In this Letter we focus on the large-scale ordering and we explain it
with phenomenological arguments supported by DNS of MHD equations modified to include expansion (expanding box model or EBM, \citealt{Grappin_al_1993,Grappin_Velli_1996}). 
The EBM has been recently used \citep{Dong_al_2014} to show the
scale-dependent competition between two axes of symmetry, the mean field axis
and the radial axis. 
Here we show that it is able to reproduce
both the large and small scale anisotropy of the SF along the three ortogonal
direction defining the frame attached to the local, scale-dependent mean field.

\section{Simulations and parameters}
We follow the evolution of a plasma volume embedded in a mean flow
with constant speed.
Turbulent evolution with distance is thus modeled as decaying, unforced
turbulence.
Two runs are analyzed: run A assumes a \textit{uniform parallel mean flow}, run B assumes a \textit{radial mean flow}, as is the solar wind.
The full MHD equations (continuity, induction, velocity and energy equations),
are integrated in time with a pseudo-spectral code on a grid of $1024^3$ points.
For run B, the MHD equations are modified to incorporate expansion, becoming
the EBM equations \citep{Dong_al_2014}. 
In the following, velocities are normalized to the initial rms
amplitude of velocity fluctuations $u_{rms}$, lengths to the box size $L^0$, and
time to the initial eddy turnover time $t_{nl}^0=L^0/2\pi u_{rms}$ . The
magnetic field $B$ is also expressed in unit of Alfv\'en speed, $B/\sqrt{4\pi\rho_0}$,
with $\rho_0$ being the average density.

\begin{figure}[t]
\begin{center}
\includegraphics [width=0.98\linewidth,clip=true]{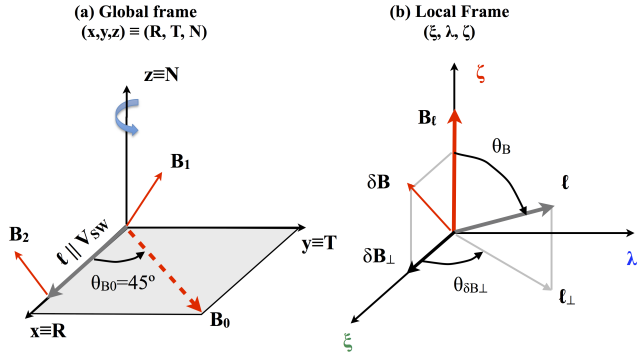}
\caption{\textit{Panel (a)}. Reference frame for simulations at $t=t^*$. The R,~T,~N directions correspond to the $x,~y,~z$ coordinates of the numerical
domain. 
The mean magnetic field $\vect{B_0}$ forms
an angle $\theta_{B0}=45^o$ with the direction of increments
$\vect{\ell}=\ell\vect{\hat{x}}$ connecting the fluctuations $\vect{B_1}$ and
$\vect{B_2}$.
\textit{Panel (b)}. Local reference frame for computing the 3D anisotropy of structure functions. 
The axes $\zeta,~\xi,~\lambda$ and the angles $\theta_B,~\theta_{\delta B\bot}$
are defined for each couple of points 
by the local magnetic field
$\vect{B_\ell}=(\vect{B_1}+\vect{B_2})/2$ and by the fluctuation
$\delta \vect{B}=\vect{B_2}-\vect{B_1}$.}
\label{fig0}
\end{center}
\end{figure}

We first define the expanding run B.
The reference frame, $x,~y,~z$, is aligned with
the R,~T,~N coordinates of the heliocentric reference frame (Figure~\ref{fig0}a).
The domain is advected by the solar wind at a
constant speed $V_{SW}$: different times $t$ correspond to different
heliocentric distances, $R(t)=R^0+V_{SW} t$, with $R^0$ being the initial
position of the simulation domain. 
During advection, the domain inflates anisotropically: the radial dimension $L_x$ does not change with time, while the lateral dimensions scale as $L_y,L_z\propto R(t)$. 
The rate of inflation is set by the expansion parameter,
$\epsilon=t_{nl}^0/t_{exp}^0=(L^0/2\pi u_{rms})/(R_0/V_{SW})$, 
where $t_{exp}^0=R^0/V_{SW}$ is the initial expansion time.
We fix the initial heliocentric distance 
$R^0=0.2~\mathrm{AU}$, the lateral dimension of the numerical domain 
$L_y^0=L_z^0=L^0=R^0/5$, and the ratio 
$u_{rms}/V_{SW}=1/4\pi$, yielding finally $\epsilon=0.4$.
The initial aspect ratio is $R_x^0=L_y^0/L_x^0=1/5$ so that at 
$R=1~\mathrm{AU}$ we have a cubic numerical domain $L_{x,y,z}=R^0=5L^0=0.2$~AU. 
The conservation of magnetic flux implies
$\vect{B_0}\propto(1/R,1,1)$,
we thus impose an oblique initial mean field,
$\vect{B_0}=[1,1/5,0]$, to have an average Parker spiral angle 
of $45^o$ at 1~AU.
Finally we set equal viscosity, resistivity, and conductivity,
$\nu=\eta=\kappa=6~10^{-5}$ 
and allow the coeffcients to vary as $1/R$ to cope with the damping of
fluctuations due to expansion. 

For the non-expanding run A ($\epsilon=0$), we choose
$L_x^0=L_y^0=L_z^0=R^0$, $\vect{B_0}=[\sqrt{2},\sqrt{2},0]$,
$\nu=\eta=\kappa=1.1~10^{-4}$.

The fluctuations $B$ and $u$ are initialized in the
same way in both runs, as a superposition of modes with random phases, with the
velocity being divergence-less.
Their spectra follow a bi-normal distribution in the Fourier space, of widths
$\sigma_\bot=4\kk^0$ and $\sigma_\|=\sigma_\bot/4$
for wavevectors perpendicular
and parallel to $\vect{B_0}$ respectively ($\kk^0=2\pi/L^0$).
The initial eddy-turnover time is thus four times smaller in the perpendicular
directions than in the parallel direction. In the expanding case, this 
reduces the expansion effects in the directions perpedicular to the radial.
The magnetic and kinetic fields are at equipartition, $B_{rms}=u_{rms}=1$,
subsonic (the sound speed is $c_s\sim7$), and have statistically vanishing correlation $\langle\vect{u}\cdot\vect{B}\rangle\sim0$ (no imbalance between the Els\"asser modes).

\begin{figure*}[t]
\begin{center}
\includegraphics [width=0.98\linewidth,clip=true]{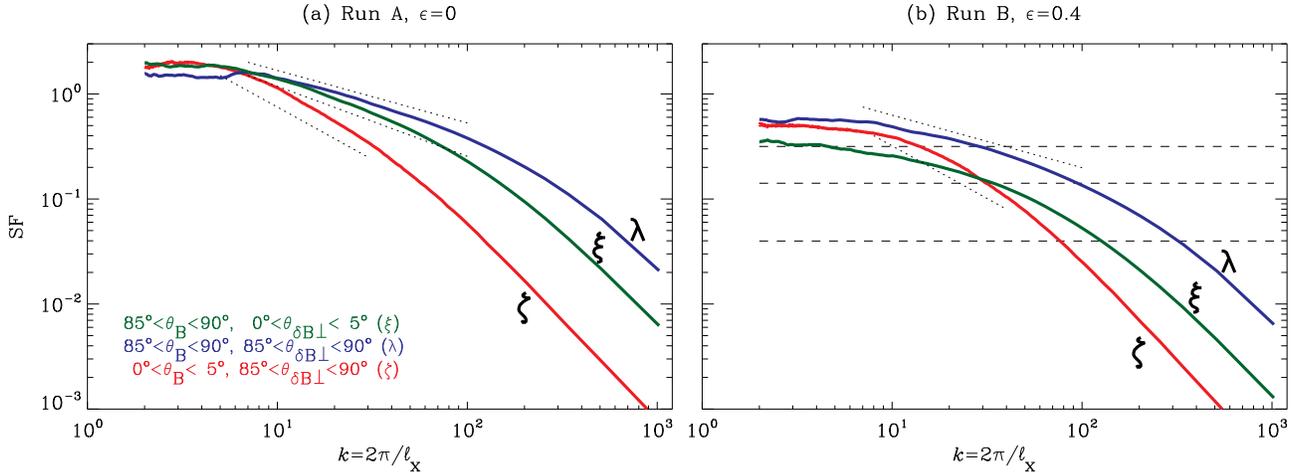}
\caption{Second order SF for the non expanding run, panel (a), and expanding
run, panel (b) (compare to Figure~1 in \citealt{Chen_al_2012}), as a function of $k=2\pi/\ell_x$. SF are accumulated at three
different couples of angles, $\theta_B$ and
$\theta_{\delta B_\bot}$, corresponding to the 
parallel direction ($\zeta$, red line) and the two perpendicular directions
($\lambda$ in blue, $\xi$ in green) respectively (see
Figure~\ref{fig0}b). 
The dotted lines are a reference for the scaling
$k^{-1/2},~k^{-2/3}$ and $k^{-1}$ in panel (a) and for $k^{-1/2},~k^{-1}$ in
panel (b). 
The horizontal dashed lines in panel (b) mark the
levels at which isosurfaces are drawn in Figure~\ref{fig2}.}
\label{fig1}
\end{center}
\end{figure*}

To compute the anisotropy of structure functions we use the procedure
described in \citet{Chen_al_2012}.
For each couple of points
$\vect{x_1},~\vect{x_2}$  separated by the increment
$\vect{\ell}=\vect{x_2}-\vect{x_1}$, we define the local mean
field as $\vect{B_\ell}=1/2(\vect{B_1}+\vect{B_2})$ and the fluctuating field as
$\delta\vect{B}=\vect{B_2}-\vect{B_1}$, where
$\vect{B_{1,2}}=\vect{B}(\vect{x_{1,2}})$.
The local scale-dependent reference frame, shown in Figure~\ref{fig0}b, has the
vertical axis $\zeta$ oriented along $\vect{B_\ell}$, the first perpendicular
axis $\xi$ oriented along the perpendicular fluctuation direction $\delta
\vect{B_\bot}\propto\vect{B_\ell}\times[\delta\vect{B}\times\vect{B_\ell}]$,
and the second perpendicular axis $\lambda$ perpendicular to both
$\vect{B_\ell}$ and $\delta\vect{B_\bot}$.
In this reference frame, the polar and azimuthal angles $\theta_B$ and
$\theta_{\delta\vect{B}\bot}$ define the direction of increment with respect
to the local mean field.
For each pair of points the $\vect{B}$-trace structure function,
$\SF=|\delta\vect{B}|^2$, is accumulated in
$5^o$ bins for $\theta_B,\theta_{\delta\vect{B}\bot}\in[0^o,90^o]$, and then
averaged in each bin (we reflected below $90^o$ any angles larger than $90^o$). 
Increments, except when otherwise stated, \textit{are computed along the $x$ direction $\vect{\ell}=\vect{\ell_x}$}, corresponding to spacecraft flybys along the radial direction in the solar wind frame, as in observations. 

We first present the results of the flyby analysis on simulated data at
$t^*=2.8$ for run A and at $t^*=10$ for run B. We then show
how the anisotropy evolves in time in the two runs.
While $t^*$ in run B is chosen to reproduce data at $R=1$~AU, the choice for
run A is arbitraty. 
Homogeneous runs evolve more rapidly
than expanding runs since, in the latter, fluctuations are damped by both
turbulence and expansion and so the nonlinear time increases more rapidly.
We thus chose a different time in run A, after the peak of current density ($t\sim2.4$)
but not too late, in order to have a Reynolds number $Re\approx1200$ and
$B_{rms}/B_0\approx0.9$ comparable to those of run B at $t^*=10$ ($B_{rms}/B_0\sim1.5$ 
and $Re\approx1300$)
\footnote{The Reynolds number is computed as $Re=(L_{inj}/L_{diss})^{4/3}$,
where $L_{inj}=(3\pi/4E)\sum_\kk E(\kk)/\kk$ and $L_{diss}=(\nu^3/D)^{1/4}$. 
$D=\sum_\kk\nu \kk^2 E(k)$ is the dissipation per unit mass and
$E(k)$ is the omnidirectional spectrum.}.

\section{Results}

In Figure~\ref{fig1}a we plot the SF of the homogenous run A as a function of
wavenumber $k=2\pi/\ell_x$ for three couples of angle $\theta_B,~\theta_{\delta
B\bot}$ corresponding to the directions $\xi,\lambda,\zeta$ in the local
reference frame of Figure~\ref{fig0}b.
Increments are taken along the $x$ direction, which forms an angle of $45^o$
with the mean field $B_0$.
At large scales $k\lesssim 8$ the SFs have comparable energy in the three directions. 
At small scales, $8\lesssim k\lesssim60$, we have $\SF(\lambda)>\SF(\xi)>\SF(\zeta)$
with the the following approximate scaling
$\SF\propto \lambda^{1/2},~\xi^{2/3},~\zeta^1$
(the power-law range is actually smaller in $\xi$
and $\zeta$).

In panel (b) we show the same plot for the expanding run B.
Its overall structure differs completely from run A. Now $\SF(\lambda)$ and
$\SF(\xi)$ have parallel profiles roughly proportional to 
$\lambda^{1/2},~\xi^{1/2}$ in the small-scale range, $10 \lesssim k
\lesssim 50$.
$\SF(\lambda)$ is dominant everywhere, while $\SF(\zeta)$ passes from almost
dominant at large scales ($k\lesssim10$) to subdominant at small scales, where the ordering is the same as for the homogeneous run A.
\begin{figure*}[t]
\begin{center}
\includegraphics [height=0.27\linewidth,clip=true]{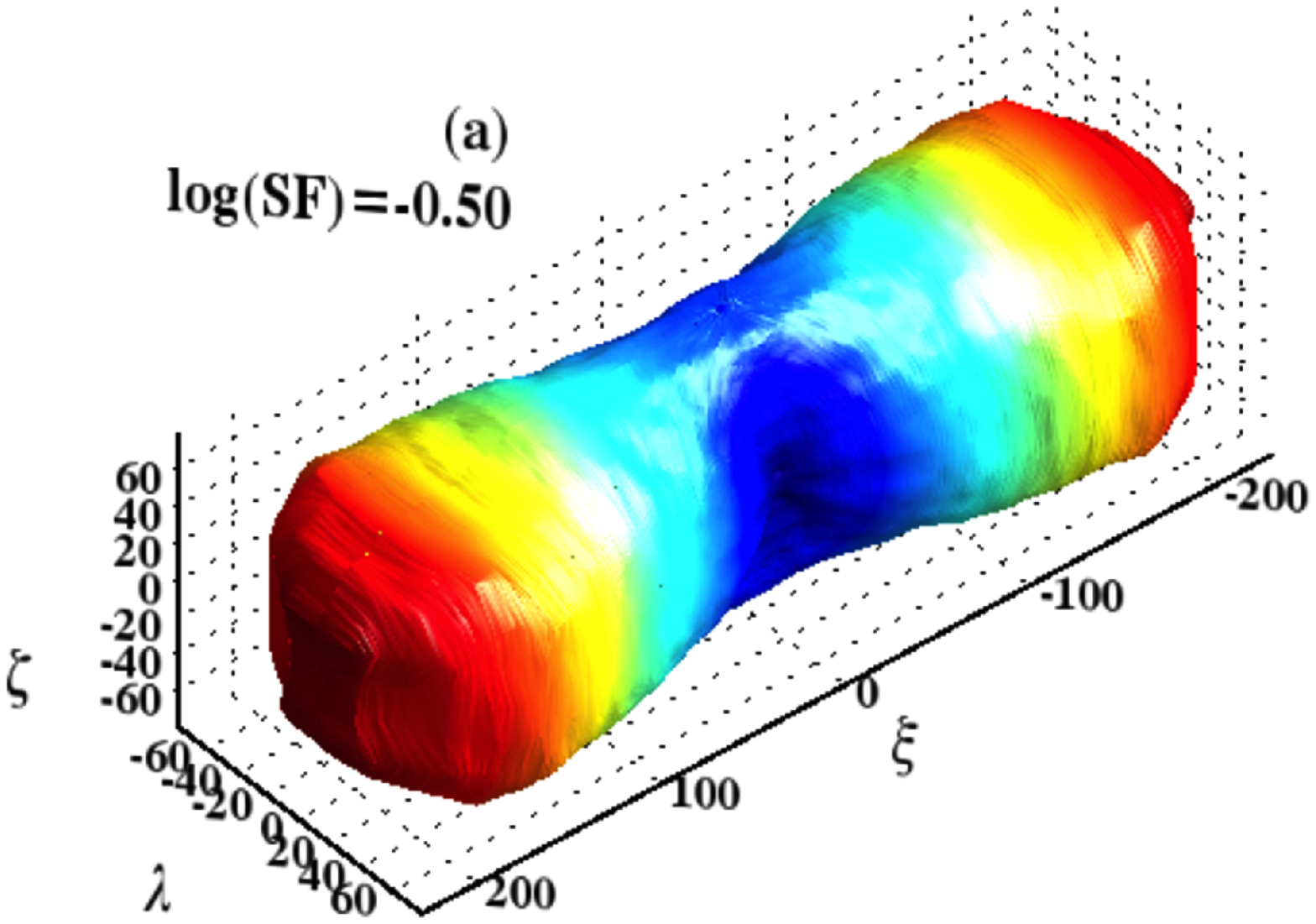}
\includegraphics [height=0.29\linewidth,clip=true]{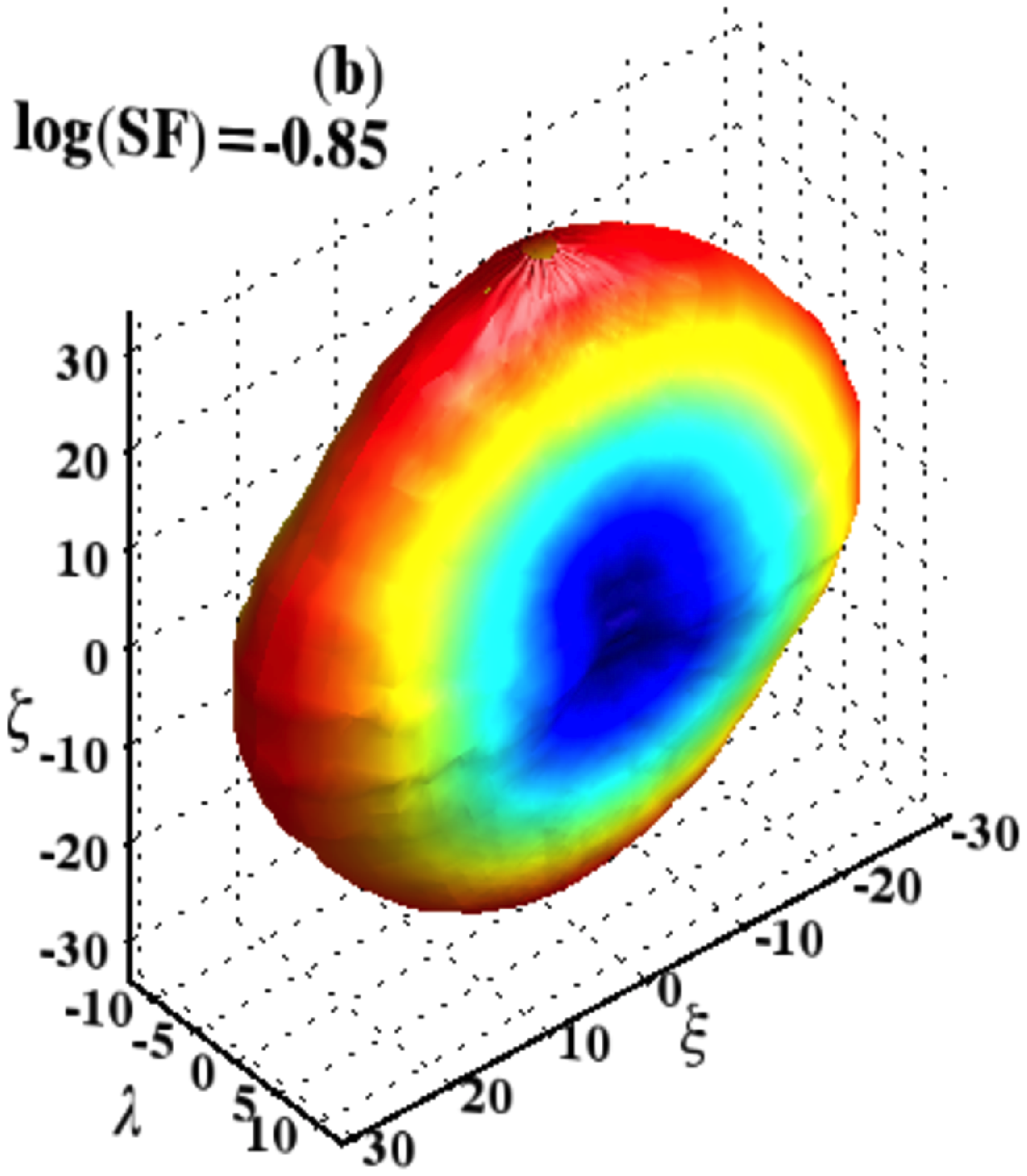}
\includegraphics [height=0.29\linewidth,clip=true]{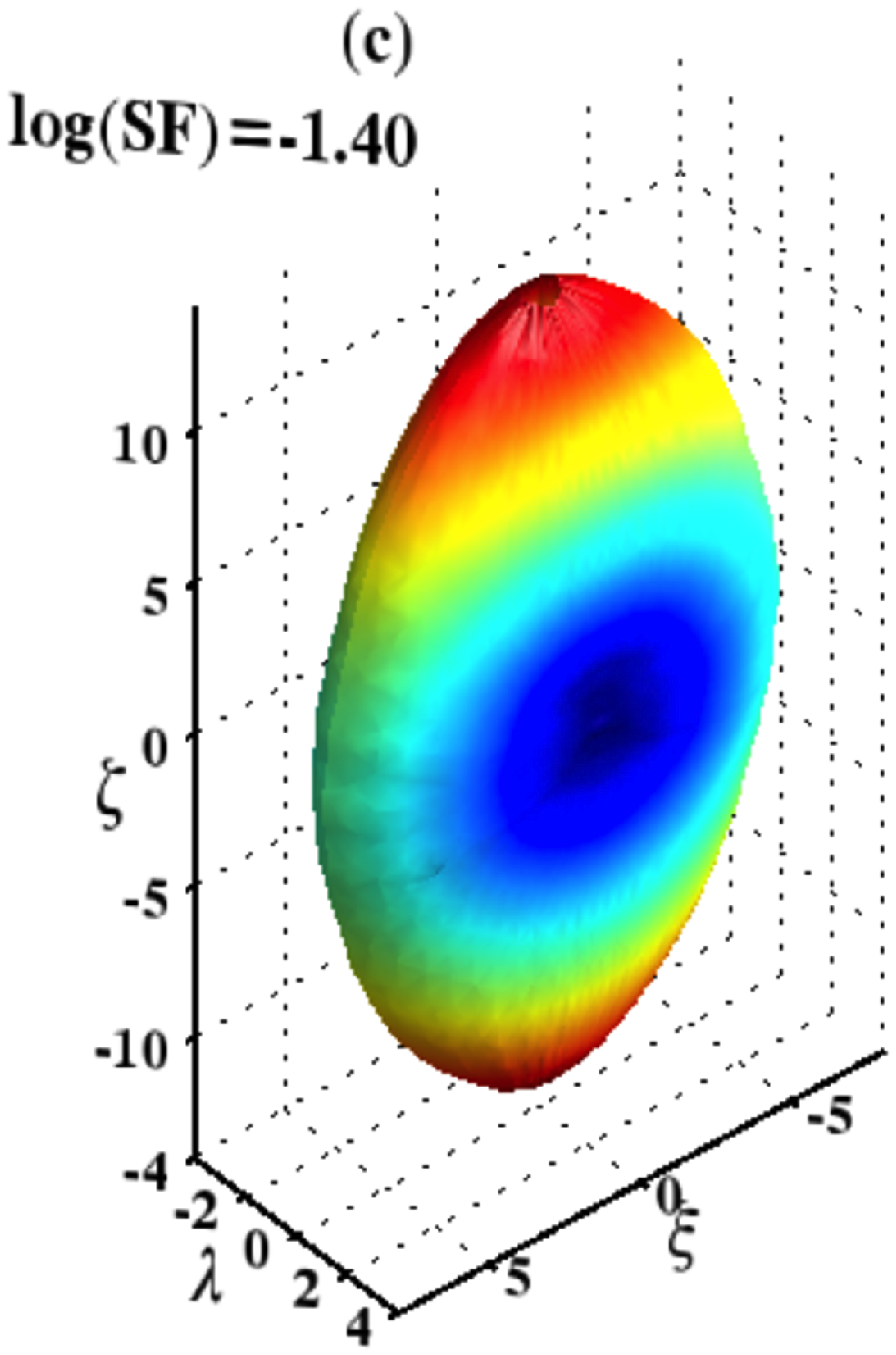}
\caption{Run B. Isosurface of constant SF power at three levels,  
$\textrm{log}\SF=-0.55,~-0.85,~-1.4$, corresponding to
smaller and smaller scales respectively (horizontal
dashed lines in Figure~\ref{fig1}).
Colors indicate distance from the origin 
to help the 3D visualization. Compare to Figure~4 in \citet{Chen_al_2012}.}
\label{fig2}
\end{center}
\end{figure*}

Following \citet{Chen_al_2012}, another viewpoint of the anisotropy in the
expanding case is given in Figure~\ref{fig2}, where we plot the isosurfaces of constant SF power at three different levels (marked as dashed lines in
Figure~\ref{fig1}b), corresponding, from left to right, to smaller and smaller scales. 
For a given value of the isosurface, its shape indicates the correlation of
fluctuations along the three directions of the local frame, and can be roughly
thought as a statistical eddy shape. Since $\SF=|\delta \vect{B}|^2$
measures the power in the anticorrelation, the more energetic is the SF along a
given direction, the smaller its correlation.
In Figure~\ref{fig2}, the smallest correlation (smallest elongation of the isosurface) is always in
the $\lambda$ direction, but the direction of the largest correlation changes with scales.
At large scales (Figure~\ref{fig2}a), the eddy is more
elongated in the $\xi$ direction, corresponding to $\delta\vect{B_\bot}$, 
at small scales (Figure~\ref{fig2}c) it becomes more elongated along the $\zeta$ direction,
corresponding to $\vect{B_\ell}$. 

The anisotropy shown in Figure~\ref{fig1}b and Figure~\ref{fig2} 
are in very good agreement with the observations of
\citet{Chen_al_2012}.
In Figure~\ref{fig1}b
the small-scale anisotropy ($10\lesssim k\lesssim50$) is roughly
consistent with critical balance, while the large-scale behavior ($k\lesssim10$) requires some more explanation.
\begin{figure*}[t]
\begin{center}
\includegraphics [width=0.48\linewidth,clip=true]{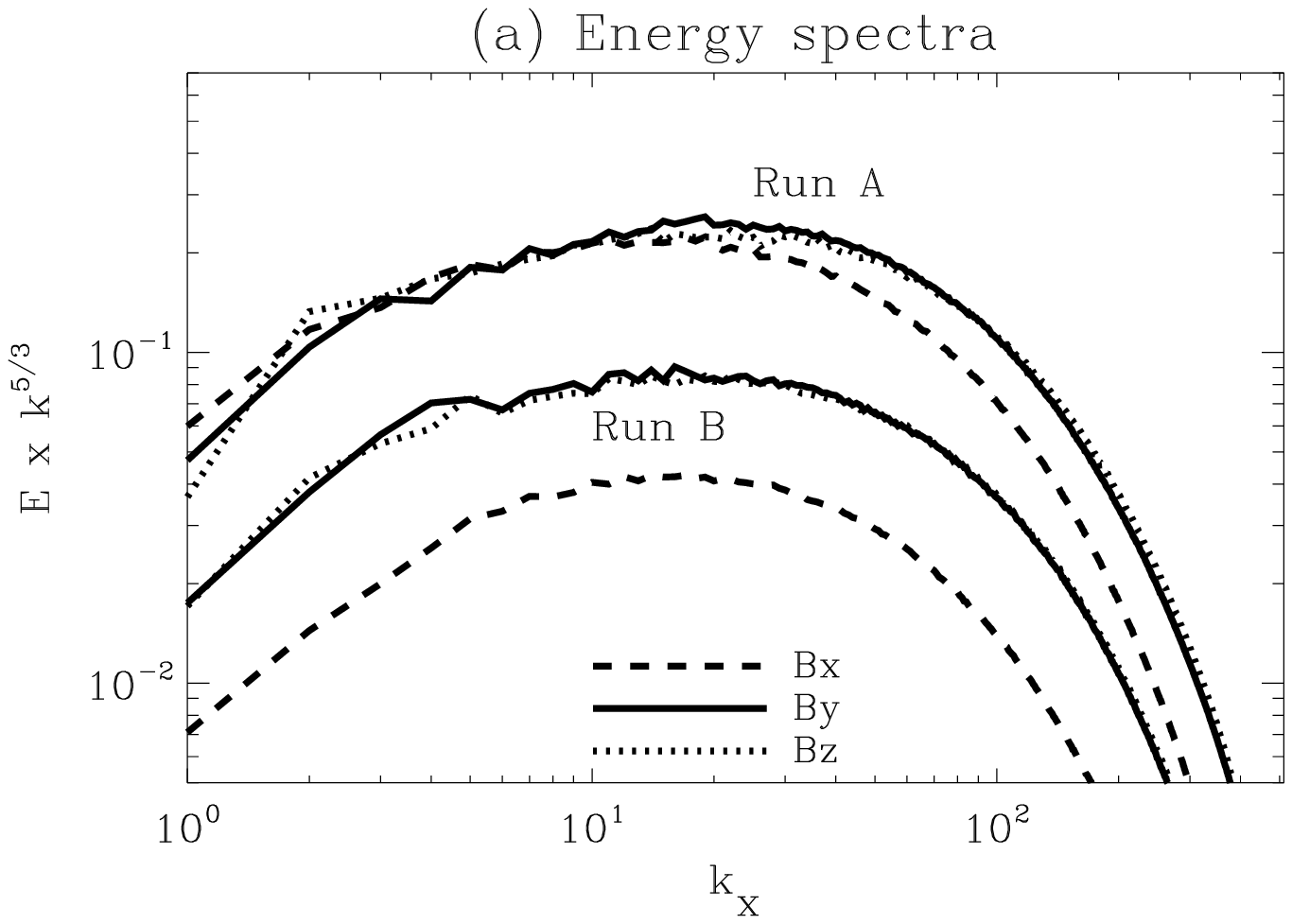}
\includegraphics [width=0.48\linewidth,clip=true]{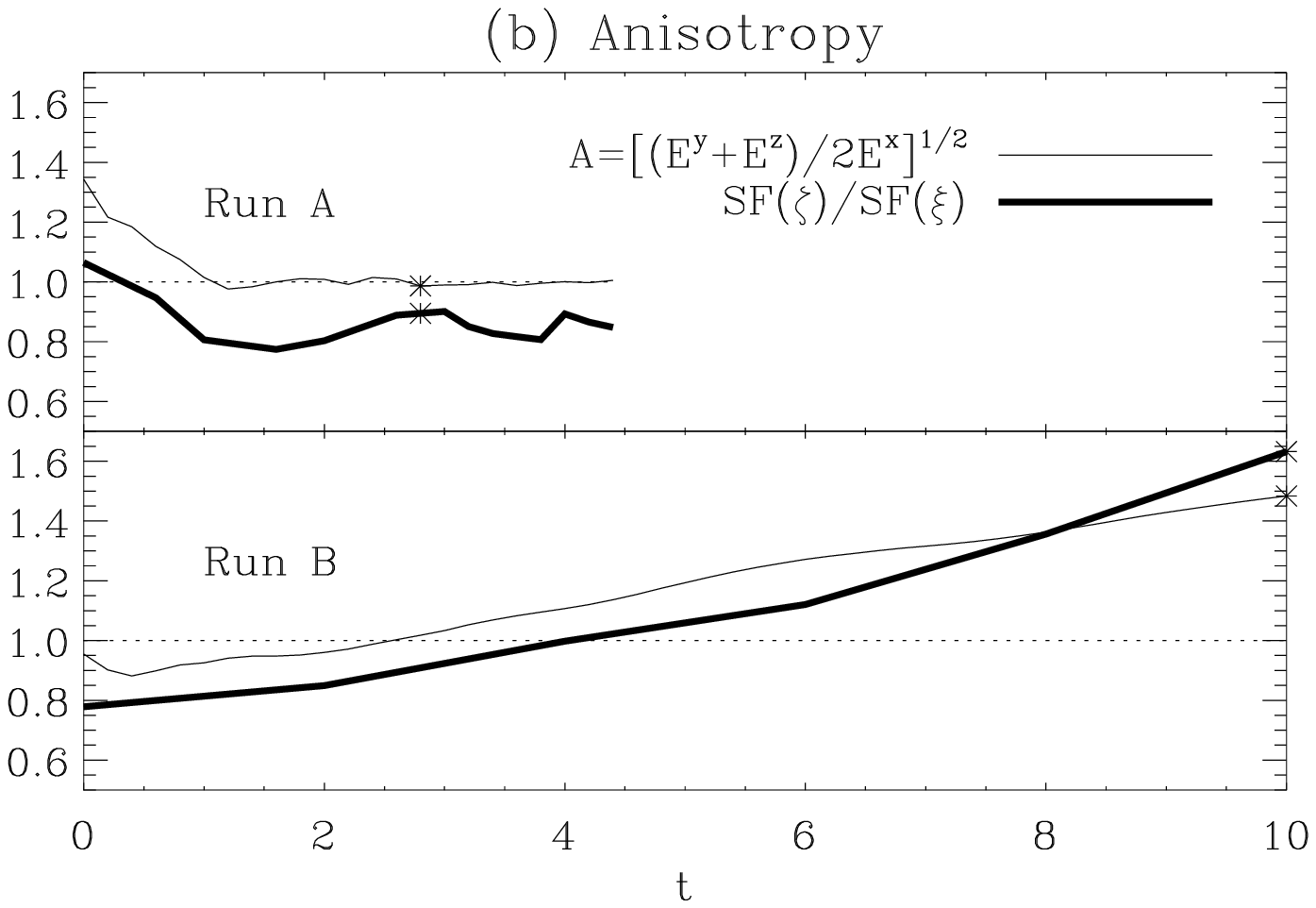}
\caption{(a) Energy spectra at $t^*$ along the radial direction $\kk_x$ for run A (top curves) and
run B (bottom curves) for the three R,~T,~N components of the magnetic field,
$B_x,~B_y,~B_z$ (dashed, solid, and dotted lines respectively).
(b) Evolution with time of the structure function anisotropy
$\SF(\zeta)/\SF(\xi)|$ (thick line) computed at
$k=2\pi/\ell_x=8$ and of the component anisotropy $A=\sqrt{(E_B^y+E_B^z)/2E_B^x}
$ (thin line) computed at $\kk_x=8$ for run A (top) and run B (bot). Asterisks
indicate the time $t^*$ at which the flyby analysis is presented.}
\label{fig3}
\end{center}
\end{figure*}

The large-scales ordering of SFs is related to the component anisotropy of
the magnetic fluctuations that originates from the selective damping induced by
the expansion.
The component anisotropy is shown in Figure~\ref{fig3}a, where we
plot the reduced energy spectra $E_B^{x,y,z}(\kk_x)$ of the
$x,~y$, and $z$ components of the magnetic fluctuations compensated by $\kk_x^{5/3}$, for runs A and B.
While in the non-expanding case energy is
distributed isotropically among the components, in the expanding case the
radial component is at least a factor 2 smaller, at all scales. 
This behavior is consistent with observations
\citep{Horbury_Balogh_2001} and is generally found in expanding runs (see
\citealt{Dong_al_2014}). 
The link between the component anisotropy and the SF
anisotropy is shown in Figure~\ref{fig3}b where we plot their evolution with
time. The SF anisotropy
is quantified as the ratio $\SF(\zeta)/\SF(\xi)$ at $k=2\pi/\ell_x=8$ (see
\citealt{Esquivel_Lazarian_2011,Burkhart_al_2014} for a similar analysis on global SF), while the component anisotropy
is evaluated as $A=\sqrt{(E_B^y+E_B^z)/2E_B^x}$ at $\kk_x=8$.
For the homogenous run A, both ratios are about constant and close to the value
of 1 (isotropy), while in run B both ratios increase steadily and approximately
with the same rate. Thus, in the expanding case as the heliocentric
distance increases the magnetic fluctuations are more
and more confined in the $y,z$ plane (the T,N plane), and so also the local
mean field will preferentially lie in this plane.

We now show that when the SF is sampled along the radial direction, the 
above component anisotropy, $E_B^y,E_B^z>E_B^x$, implies that the SF has
a different power along the three directions $\xi,\lambda,\zeta$ defining the
local reference frame.
Consider two vectors $\vect{B_1},~\vect{B_2}$ at positions $x_1$ and $x_2$ and indicate with 
$\vect{B^\bot_{1,2}}$ their components in the $y,z$
plane ($\vect{B^\bot}=\vect{B_y}+\vect{B_z}$), with $\alpha$ the angle between
them, and with $B^\|_{1,2}$ their projection along $x$.
Assume also for simplicity 
\begin{eqnarray}
B_{1,2}^\|\sim\ord{2}<<B^\bot_1=\ord{1}=1.
\label{eq1}
\end{eqnarray}
The local mean field and the
fluctuation in the $y,z$ plane are given by
\begin{eqnarray}
\vect{B}_\vect{\ell}^{\bot}=\sqrt{1+{B^\bot_2}^2+2B^\bot_2\cos\alpha},
\label{eq2}\\
\delta\vect{B}^{\bot}=\sqrt{1+{B^\bot_2}^2-2B^\bot_2\cos\alpha}
\label{eq3}
\end{eqnarray}
($\delta \vect{B^\bot}$ should not be confused with $\delta \vect{B_\bot}$ 
that defines the local reference frame). 
These equations simply state that
when $\vect{B}_1^\bot$ and $\vect{B}_2^\bot$ are aligned (anti-aligned) the
local field $B_\ell^\bot$ is large (small) and the fluctuating field 
$\delta B^\bot$ is small (large).
The orientation of the local reference frame with respect to the fixed
radial direction $x$ determines which local SF we are
measuring: we cumulate the power in $\SF(\lambda)$ or in $\SF(\zeta)$ or in
$\SF(\xi)$ when $\lambda$ or $\zeta$ or $\xi$ lies along $x$.
We thus estimate the leading order contributions associated to each of them
by considering the power $|\delta\vect{B}|^2$ associated to the above three
orientations:
\begin{itemize}
\item To have power in $\SF(\lambda)$ one needs both $\vect{B_\ell}$ and
$\delta\vect{B}$ to lie in the $y,z$ plane, that is,
$B_\ell^\bot>>B_\ell^\|$ and $\delta B^\bot>>\delta B^\|$. This condition is
readily satisfied from Equation~\eqref{eq1}, provided that
$B^\bot_2<<B^\bot_1$, or $B^\bot_2\sim B^\bot_1$ and
$60^o\lesssim\alpha\lesssim 120^o$ in Equations~\eqref{eq2}-\eqref{eq3},
yielding in both cases 
$\SF(\lambda)\sim|\delta\vect{B^\bot}|^2\sim\ord{1}$.
\item The contribution to $\SF(\zeta)$ is obtained when $\vect{B_\ell}$ is aligned
along $x$, that is $B_\ell^\bot<<B_\ell^\|$. This happens only
when $B^\bot_2\sim B^\bot_1$ and $120^o<<\alpha\lesssim180^o$ in
Equation~\eqref{eq2}, implying from Equation~\eqref{eq3} a contribution
$\SF(\zeta)\sim|\delta\vect{B^\bot}|^2\sim\ord{1}$.
\item Finally, to have power in $\SF(\xi)$, $\vect{B_\ell}$ must belong 
the $y,z$ plane and $\delta\vect{B}$ must have a non negligible $x$
component $\delta B^\|>\delta B^\bot/10$ (our minimum angular resolution is $5^o$).
The conditions $\delta B^\bot\sim\delta
B^\|\sim\ord{2}$ and $B_\ell^\|<<B_\ell^\bot\sim\ord{1}$ are satisfied in
Equations~\eqref{eq2}-\eqref{eq3} only when $B^\bot_2\sim B^\bot_1$ and
$\alpha<<60^o$, which yield a power
$\SF(\xi)\sim|\delta\vect{B^\bot}|^2\sim|\delta\vect{B^\|}|^2\sim\ord{2}$.
\end{itemize}
Thus, the geometrical constraint imposed by the component anisotropy induced by
expansion, $E_B^y,E_B^z>E_B^x$, favors a local anisotropy with
$\SF(\lambda)\ge\SF(\zeta)>\SF(\xi)$,
as is indeed found at large scales in Figure~\ref{fig1}b and in the
solar wind observations. 
\begin{figure*}[t]
\begin{center}
\includegraphics [width=0.98\linewidth,clip=true]{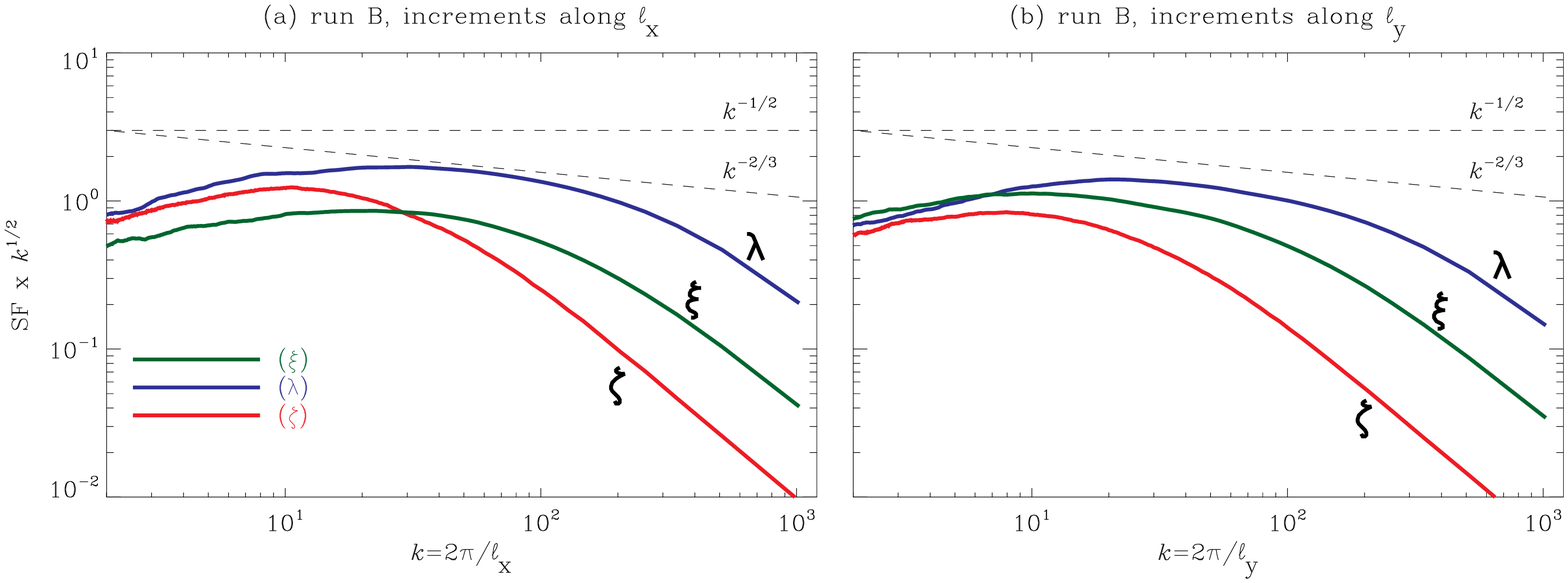}
\caption{SF of expanding run B compensated by $k^{1/2}$ 
as a function of increments computed along
the radial directions $k=2\pi/\ell_x$ (left), and along the transverse
direction $k=2\pi/\ell_y$ (right). The two dashed lines are a reference for the
scaling $k^{-1/2}$ and $k^{-2/3}$.}
\label{fig4}
\end{center}
\end{figure*}

We finally show how the anisotropy of expanding turbulence changes if
one samples increments in directions perpendicular to the radial.
In Figure~\ref{fig4} we plot the SFs of run B computed along $\ell_x$ and
$\ell_y$, corresponding to the R and T directions, respectively.
The SFs are now compensated by $k^{1/2}$ to highlight 
inertial-range scales and the corresponding spectral index.
Independently of the direction of increments, the inertial range extends to smaller scales in the second perpendicular
direction ($\lambda$) than in the first perpendicular direction ($\xi$).
The parallel direction ($\zeta$) does not show any convincing scaling, although having a steeper spectrum.
The direction of increments affect the overall anisotropy.
In fact, the large-scale ordering, characteristic of expansion in panel (a),
disappears when increments are along the transverse direction (panel (b)), the
SF becoming basically isotropic for $k\lesssim10$. 
This can be interpreted as a reduced effect of the component
anisotropy when increments are along the transverse direction (a similar
behavior is seen for $\ell=\ell_z$, not shown).
Moreover the two perpendicular SFs (blue and green curve) exhibit a
different spectral index, passing from a slope $\approx 1/2$ in panel (a) to
steeper spectral index $\approx 2/3$ in panel (b).
Note finally that the eddy shape for transverse increments is at all scales
qualitatively similar to the anisotropy in Figure~\ref{fig2}c, and thus differs
completely from the cases of radial increments shown in Figure~\ref{fig2}a,b.

\section{Discussion}
We computed the 3D anisotropy of structure functions with respect to the local mean field in DNS of MHD turbulence, including or not expansion.
For homogenous turbulence, the SF is roughly isotropic at large scales and develops scale-dependent anisotropy at small scales 
due to the different scaling along the different local directions.
The corresponding spectral indices are roughly consistent with $-1/2,~-2/3,$ and $-1$ in the $\lambda,~\xi$ and $\zeta$ directions respectively.
Such an ordering has been predicted by
\citet{Boldyrev_2006} and implies that $\SF(\lambda)>\SF(\xi)>\SF(\zeta)$ at
small scales. The homogenous run qualitatively agrees with
the above power anisotropy, which is stable although the precise slopes vary
with time and with sampling direction.

When expansion is taken into account the anisotropy is determined by a
well-defined large-scale power anisotropy and by a different scaling along the
parallel direction ($\zeta$) and the two perpendicular directions
($\lambda,~\xi$), the latter now having approximately the same spectral index ($\approx -1/2$).
We found that the overall SF anisotropy is a consequence of the component
anisotropy induced by expansion and that it shows up only when increments are
computed along the radial direction.
When increments are along the transverse
direction, the large-scale anisotropy disappears, the eddy shape does not change with scales
(although the anisotropy increases at smaller and smaller scales),
and the SFs exhibit steeper spectral scaling.
Thus, the measured anisotropy of solar wind turbulence would change for increments in 
directions other than the radial, a situation that may become possible to
test with Solar Probe Plus in its near-sun orbital phase.

Let us compare our results with the observations 
in the fast solar wind \citep{Chen_al_2012}.
In run B the choice of the solar wind speed yields,
via the Taylor hypothesis, the spacecraft frequencies corresponding to the
radial increments in Figure~\ref{fig1} and Figure~\ref{fig2}.
For a fast wind $V_{SW}\sim800~\mathrm{km~s^{-1}}$, one gets
$f_{min}=V_{SW}R_x^0/L^0\sim 3~10^{-5}~\mathrm{Hz}$ and
$f_{max}=512\times f_{min}\sim1.5~10^{-2}~\mathrm{Hz}$.
Our initial fluctuations have vanishing $u,B$ correlations for easier
comparison between the expanding and homogenous runs, while fast wind has high correlations.
Contrary to the non-expanding case, initially highly correlated
fluctuations lead to fully developed turbulence in
expanding runs \citep{Verdini_Grappin_2015},
with a SF anisotropy similar to (expanding) runs with vanishing
$u,B$ correlations.
Finally, for scales in between 5-10 hours, the ratio
$u_{rms}/V_{SW}\approx0.1$ for both fast and slow wind \citep{Grappin_al_1990}, 
thus the expanding parameter $\epsilon=0.4$ is suited for both types of
wind.
To conclude, as far as the local anisotropy is concerned, the results presented in this Letter are representative of both fast and slow wind
and match observations in the fast solar wind, including the change of
anisotropy with scales.
They indicate that expansion, by distributing energy among different components
of the magnetic fluctuations, affects the local mean field orientation and
hence the observed anisotropy with respect to it.
The present results complement those ones obtained in the recent numerical study
by \citet{Dong_al_2014} on the \textit{global anisotropy} in the solar wind. Both studies
show that expansion strongly affects anisotropy of solar wind turbulence at
inertial range scales. Finally, the convergence found with observations proves 
the validity of the EBM approach to model and study solar wind turbulence.\\

\textit{Acknowledgments} 
We thank the referee for useful and constructive comments. 
This project has received funding from the European Union's Seventh
Framework Programme for research, technological development and demonstration
under grant agreement No. 284515 (SHOCK). Website: project-shock.eu/home/. AV
acknowledges the Interuniversity Attraction Poles Programme initiated by the Belgian Science Policy Office (IAP P7/08 CHARM).  
HPC resources were provided by CINECA (grant 2014 HP10CLF0ZB and
HP10CNMQX2M). 

\bibliographystyle{apj}

\end{document}